# Revisiting the Reduction of Thermal Conductivity in Nano- to Micro-Grained Bismuth Telluride: The Importance of Grain-Boundary Thermal Resistance

Sien Wang,[1†] Xiaowei Lu,[2†] Ankit Negi,[3] Jixiong He,[3] Kyunghoon Kim,[3] Hezhu Shao,[4] Peng Jiang,[2] Jun Liu,[3*] and Qing Hao[1*]

[1] Department of Aerospace and Mechanical Engineering, University of Arizona, Tucson, AZ 85721, United States

[2] State Key Laboratory of Catalysis, CAS Center for Excellence in Nanoscience, Dalian Institute of Chemical Physics, Chinese Academy of Sciences, Dalian 116023, China

[3] Department of Mechanical and Aerospace Engineering, North Carolina State University, Raleigh, NC 27695, United States

[4] College of Electrical and Electronic Engineering, Wenzhou University, Wenzhou 325035, China

Abstract

Nanograined bulk alloys based on bismuth telluride ($Bi_2Te_3$) are the dominant materials for room-temperature thermoelectric applications. In numerous studies, existing bulk phonon mean free path (MFP) spectra predicted by atomistic simulations suggest sub-100 nm grain sizes are necessary to reduce the lattice thermal conductivity by decreasing phonon MFPs. This is in contrast with available experimental data, where a remarkable thermal conductivity reduction is observed even for micro-grained $Bi_2Te_3$ samples. In this work, first-principles phonon MFPs along both the in-plane and cross-plane directions are re-computed for bulk $Bi_2Te_3$. These phonon MFPs can explain new and existing experimental data on flake-like $Bi_2Te_3$ nanostructures with various thicknesses. For polycrystalline $Bi_2Te_3$-based materials, a better explanation of the experimental data requires

---

further consideration of the grain-boundary thermal resistance that can largely suppress the transport of high-frequency optical phonons.

Key words: Thermal conductivity, phonon mean free path, grain boundary, first principles calculations, thermal resistance.

In the last two decades, nanostructured bulk (nano-bulk) materials have been widely studied as an effective approach to improve the thermoelectric performance of materials such as $Bi_2Te_3$-based alloys[1-3] and Si,[4, 5] which benefits from a significantly reduced lattice thermal conductivity ($k_L$) and maintained bulk-like electrical properties. These nano-bulk materials are prepared by grinding down the bulk ingot or mixed elemental chunks into nanopowder and then hot pressing the nanopowder into a bulk disc with a high volumetric density of grain boundaries (GBs) inside.[1, 3] Despite many experimental results, a good understanding of the observed $k_L$ reduction is still lacking. Due to the significant grain growth during the hot press, the observed grain sizes in the nano-bulk samples are widely distributed from ~10 nm to ~1 μm, where the averaged grain size can be above 1 μm.[2] In contrast, first-principles or molecular dynamics (MD) simulations often suggest the bulk phonon mean free paths (MFPs) in $Bi_2Te_3$ are below 100 nm at room temperature.[6-8] In this situation, micro-grains should only slightly reduce the $k_L$ based on classical phonon size effects, in which the $k_L$ reduction is solely attributed to the reduced effective phonon MFP ($\Lambda_{eff}$) modified from the bulk phonon MFP ($\Lambda_{bulk}$) according to the Matthiessen's rule, i.e., $1/\Lambda_{eff} = 1/\Lambda_{bulk} + 1/L_C$.[9] For polycrystalline materials, the characteristic length is $L_C = \alpha d$ ($0 < \alpha \leq 1$), with $d$ as the grain size and $\alpha = 1$ in a simple treatment.[10] Although good agreements between theoretical modeling and measurements were obtained for some micro-grained $Bi_2Te_3$ samples, the employed $\Lambda_{bulk}$ was determined by fitting the bulk $k_L$ for $Bi_2Te_3$ and was up to ~ 1 mm at 300 K,[11-13] which was in big contrast with predictions by atomistic simulations.[6-8] Similar puzzles were also found in wurtzite ZnO, where the first-principles bulk phonon MFPs cannot explain the observed $k_L$ reduction in polycrystalline ZnO thin films and the divergence was only ascribed to defects in real films.[14] In this aspect, the discrepancy between computed phonon MFPs and measured $k_L$ reduction in polycrystalline materials can be eliminated by considering the interfacial thermal resistance or Kapitza resistance ($R_K$) at GBs,[15] which results from the frequency-dependent phonon transmission or reflection at GBs.[16, 17] For nano-grained bulk Si, low-temperature $k_L$ fitting required a frequency-dependent $L_C \sim \alpha d/\omega$ due to the existence of $R_K$, with $\omega$ as the angular frequency of phonons.[10] In a more advanced effective medium formulation (EMF), the GB phonon transmissivity $T_{GB}$ and further the probability $P_{GB}$ for specular GB phonon transmission can be individually treated, in addition to the $\Lambda_{eff}$ reduced from $\Lambda_{bulk}$.[18-20] This EMF eliminates the fitted scaling factor $\alpha$ for $L_C$ when $T_{GB}$ and $P_{GB}$ can be predicted by atomistic simulations[21-23] and machine learning models.[24]

For polycrystalline $Bi_2Te_3$, the phonon transport analysis should also consider the thermal anisotropy of $Bi_2Te_3$. Figure 1a shows the crystal structure of $Bi_2Te_3$, where one period as five layers are repeated along the *c*-axis or the cross-plane direction: $Te_1$-Bi-$Te_2$-Bi-$Te_1$. Between adjacent periods, $Te_1$ atoms have van der Waals interactions. Discrepancies can often be found among existing computational and experimental data of the anisotropic $k_L$ for single-crystal $Bi_2Te_3$. In very early experimental studies, the room-temperature $k_{L,\parallel}$ was found to be 1.30,[25] 1.54,[26] 1.70,[27] 1.72,[28] and 1.77 W/m·K[29] in different measurements. For samples with highly oriented large grains, $k_{L,\parallel}$ was roughly measured as 1.65 W/m·K at 300 K.[30] In comparison, recent theoretical predictions suggest $k_{L,\parallel} \approx$ 1.73 W/m·K[31] and $k_{L,\parallel} \approx$ 1.27 W/m·K.[8] Along the cross-plane direction, $k_{L,\perp} \approx$ 0.64 – 0.79 W/m·K at 300 K was measured by Goldsmid,[32] whereas $k_{L,\perp} \approx$ 0.64 W/m·K was suggested by Jacquot et al.[31] and a lower $k_{L,\perp} \approx$ 0.37 W/m·K was obtained from first-principles calculations by Hellman and Broido.[8] Above divergence may result from the measurement techniques, defects in real samples, and parameters/assumptions employed in different atomistic simulations. One way to validate the computed $k_L$ and its phonon MFP spectra is by measuring the thermal conductivity of single-crystal nanostructures with varied sample sizes and thus $L_C$, as demonstrated in thermal studies for Si nanowires[33, 34] and Si thin films.[35, 36] For anisotropic $Bi_2Te_3$, such measurements must be along the major axis directions as well. Accurate phonon MFP spectra are essential for the data analysis of polycrystalline samples to evaluate the $T_{GB}$ and thus $R_K$ influence at GBs.

In this work, single-crystal $Bi_2Te_3$ nanoflakes of various thicknesses (~20 nm to 300 nm) were synthesized with chemical vapor deposition (CVD)[37] and measured for their room-temperature $k_{L,\perp}$ values using the time-domain thermoreflectance (TDTR) method. The measurement results are used to verify the first-principles accumulated $k_{L,\perp}$ re-computed for bulk $Bi_2Te_3$. In addition, the accumulated $k_{L,\parallel}$ is also computed and is used to re-analyze existing experimental results for $Bi_2Te_3$ nanostructures. Based on these anisotropic bulk phonon MFPs, EMF predictions with further directional averaging for the $k_L$ are compared with measurement data for nano- and micro-grained $Bi_2Te_3$-based bulk samples[3, 38-43] or thin films.[12, 13] It is found that optical phonons with higher frequencies and thus largely decreased GB phonon transmissivities should be strongly suppressed for their $k_L$ contributions to explain the divergence between the measurements and predictions at microscale grain sizes. For $Bi_2Te_3$-based alloys, a better agreement between experiments and EMF predictions is obtained using the direction-averaged acoustic-phonon MFPs fitted for a $Bi_{0.5}Sb_{1.5}Te_3$ bulk alloy with 20-μm grain sizes,[44] as point-defect phonon scattering in alloys is further addressed. In EMF calculations, an averaged $T_{GB}=0.5$ for phonons is found to be reasonable for the data analysis, whereas $\Lambda_{eff}$ is only slightly reduced from $\Lambda_{bulk}$ for sub-1 μm or larger grains to impact the $k_L$. This indicates the importance of $R_K$ in the data analysis of general polycrystalline samples. Based on the full phonon dispersion and the diffuse mismatch model (DMM) suggesting $T_{GB}=0.5$, the $R_K$ value is estimated to be around $6.8 \times 10^{-8}$ K·m²/W at room temperature, which is consistent with $R_K \sim 10^{-9} - 10^{-8}$ K·m²/W measured.[45, 46] If all optical phonons are filtered out by GBs, the DMM-predicted resistance will be increased to $2 \times 10^{-7}$ K·m$^2$/W. Along this line, more studies are thus required to engineer the GBs to maximize the thermoelectric performance of these materials.[40, 47-51]

*Materials synthesis and sample preparation.* Bi$_2$Te$_3$ nanoplates with different lateral sizes and thicknesses were synthesized by the well-established vapor−solid (VS) mechanism (see Supporting Information for details).[52, 53] A scanning electron microscope (SEM) was employed to characterize the morphological and compositional information on the Bi$_2$Te$_3$ nanoplates (Figure 1b). The lateral size of nanoplates with a uniform composition can be up to ~50 μm. The typical sample thickness ranges from ~10 nm to hundreds of nanometers, as revealed by an atomic force microscope (AFM), as shown in Figure 1c. The crystal quality was characterized by the micro-Raman spectroscopy, where two vibrational modes of $E_g^2$ (101 cm$^{-1}$) and $A_{1g}^2$ (131.9 cm$^{-1}$) can be clearly recognized (Figure 1d). The high crystallinity of all nanoflakes is indicated by the uniform distribution for the peak intensity for the $E_g^2$ mode.

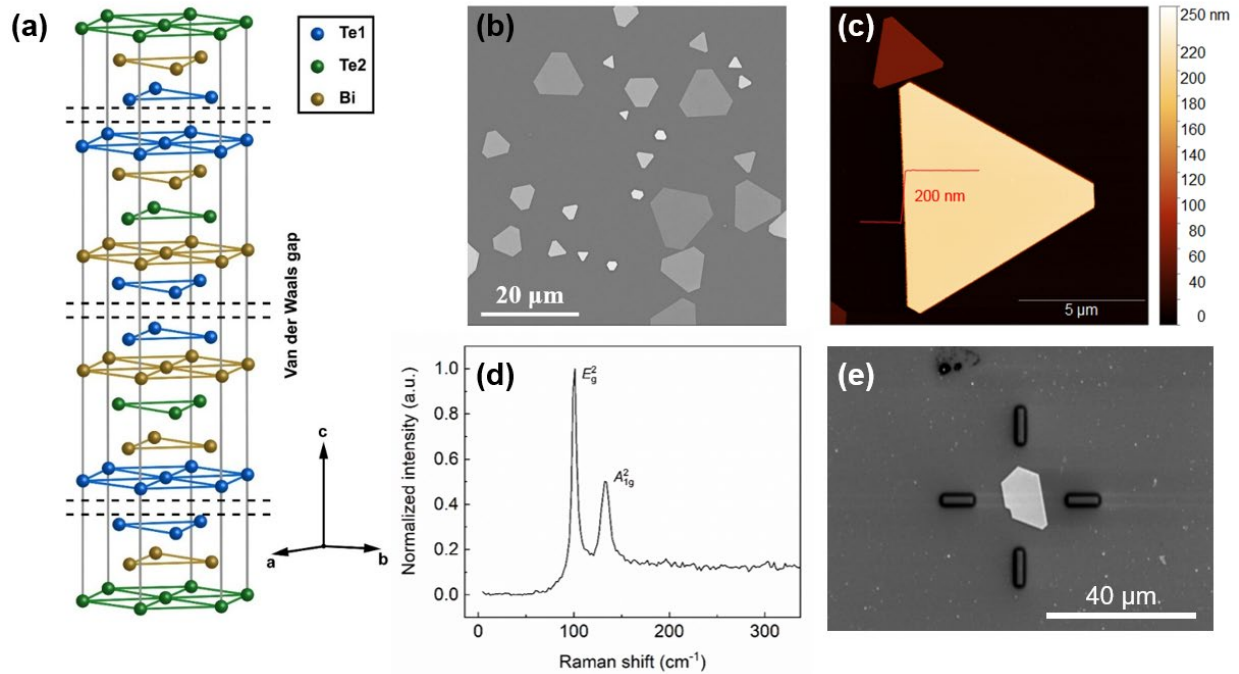

Figure 1. (a) Crystal structure of $Bi_2Te_3$. (b) SEM image of $Bi_2Te_3$ nanoplates on a $SiO_2$/Si substrate. (c) A typical AFM scan to check the sample thickness. (d) Raman spectroscopy of a typical sample with clear $E_g^2$ and $A_{1g}^2$ modes. (e) SEM image of the 63-nm-thick $Bi_2Te_3$ flake transferred onto a new $SiO_2$/Si substrate, with Al deposited across the surface.

Due to the high density of samples on their initial substrate, samples were dry-transferred onto a new clean $SiO_2$ substrate using a thermal release tape (TRT),[46, 54, 55] as detailed in the Supporting Information. Most samples were able to remain intact and wrinkle-free after the transfer. AFM studies show an undulation of less than 3 nm for the transferred sample surface. After thickness checking with an AFM, the $Bi_2Te_3$ nanoplates were then deposited with a ~80-nm-thick Al layer as the transducer for TDTR measurements. Figure 1e shows a typical sample with fabricated markers.

*TDTR measurements.* At room temperature, the cross-plane $k_{L,\perp}$ of selected samples was measured using the TDTR method (see measurement details in Supporting Information). TDTR is an ultrafast laser-based, accurate, and robust technique that has been employed to probe various thermal properties, including thermal conductivity, interfacial thermal conductance, and heat capacity of sample systems ranging from thin films, bulk substrates, to nanoparticles.[56-58] Figure 2 shows the thickness-dependent $k_{L,\perp}$ of $Bi_2Te_3$ flakes or thin films, and its comparison with literature data and theoretical predictions. Details for the theoretical predictions are also included in the Supporting Information. The lattice thermal conductivity $k_{L,\perp}$ is calculated by subtracting the estimated electronic contribution $k_E$ (~10% of the total $k_\perp$, see Supporting Information) from the measured total cross-plane thermal conductivity by TDTR. To understand and validate the measured thickness-dependent $k_{L,\perp}$ and other existing data, we performed first-principles predictions of the phonon mode-resolved thermal conductivity (with details described in the next section) and then converted this modal thermal conductivity to the film-thickness-dependent thermal conductivity using suppression functions for the in-plane[59] and cross-plane[60] directions. We found that the measured lattice thermal conductivity $k_{L,\perp}$ matches well with our theoretical predictions. The predictions in $k_{L,\parallel}$ also matches with the general trend of the diverse literature data.

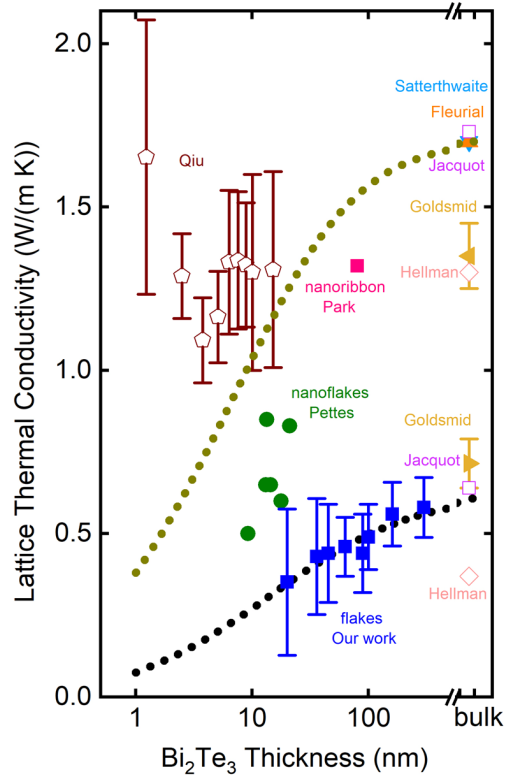

Figure 2. Room-temperature $k_{L,\perp}$ (blue filled squares as experimental data) as a function of $Bi_2Te_3$ flake thickness $t$. The measured $k_{L,\perp}$ from Goldsmid[32] is shown as the yellow triangle. In comparison, first-principles calculations from our work (black dotted line) and Hellman[8] (pink diamond) are also shown. Theoretical prediction of $k_{L,\perp}$ from Jacquot[31] is shown as the pink open square. Literature measurement results for room-temperature $k_{L,\parallel}$ from Pettes[61] (green filled circles) on a nanoflake, from Park[62] (red square) on a nanoribbon, from Goldsmid[32] (yellow triangle) on a crystal, from Fleurial[63] (orange triangle) and Satterthwaite[64] (blue triangle) on single crystals are shown. First-principles calculations from our work (dark yellow dotted line) and Hellman[8] (pink diamond), and molecular dynamics calculations from Qiu[65] (wine pentagon) are also shown. Suppression functions in the in-plane[59] and cross-plane[60] directions have been considered in our first-principles results by converting from phonon-MFP-dependent thermal conductivity to film-thickness-dependent thermal conductivity.

*First-principles calculations.* To justify the phonon size effects, the modal phonon MFPs and accumulated $k_{L,\perp}$ and $k_{L,\parallel}$ were computed with first-principles calculations. Fully anharmonic *ab initio* first-principles calculations were performed based on density functional theory (DFT), as implemented by the Vienna Ab initio Simulation Package (VASP).[66, 67] The stochastically initialized temperature-dependent effective potential method (s-TDEP) method[68, 69] was used to optimize the force constants. The projector-augmented-wave (PAW) potential and local-density approximation (LDA) were employed for the DFT calculations.[70] The LDA was applied to compute the electron density distribution iteratively where we do not need to explicitly consider different types of interlayer interactions, such as the van der Waals (vdW) or Coulomb interactions. The Bi ($5d^{10}6s^26p^3$) and Te ($5s^25p^4$) were treated as valence states. For the calculations of thermal conductivities of $Bi_2Te_3$, the values, especially the out-of-plane thermal conductivities, are extremely sensitive to the volumes. We found that the lattice parameters significantly deviate from the experimental values if the vdW functions instead of LDA were used. For example, vdW-DF and vdW-DF2 consistently result in a much lower bulk modulus than the LDA results, then underestimate the phonon frequencies. This finding is consistent with the work by Hellman et al.,[8] where the LDA method describes the experimental phonon dispersion well but the vdW functions underestimate the phonon frequencies. The reported experimental crystal parameters were collected in the AFLOW library,[71, 72] total 13 groups so far, of *a* and the angle *alpha* ranging from 10.4455 Å to 10.4784 Å, and 24.14° to 24.25682°, respectively. We then averaged these values of

*a* and *alpha* (*a* of 10.4675 Å and *alpha* of 24.1878°) and set them as the initial lattice parameters and then used LDA for structural relaxation. The energy cutoff for plane-wave, the electronic energy convergence, and the force convergence were set to be 350 eV, $10^{-7}$ eV, and $10^{-3}$ eV/ Å, respectively, to ensure sufficient ion relaxations. To fully consider the phonon renormalization effects at room temperature, we employed the s-TDEP method to calculate the force constants and thermal conductivity of $Bi_2Te_3$. Firstly, we obtained the initial interatomic force constants (IFCs) from pair potentials together with Debye model and the atomic reconfiguration was allowed to relax at room temperature. The new IFCs were fitted from the force-displacement dataset. Then the new IFCs were used to generate new stochastic configurations and iterated until convergence. We used 50 configurations during each round to obtain the IFCs, and the results were typically converged after four iterations. Every configuration contains 96 Bi atoms and 144 Te atoms, which is a 4×4×1 supercell of hexagonal unit cell of $Bi_2Te_3$. A 0.759 nm force interaction cutoff distance was found to converge the second-order and third-order force interactions. After obtaining the second-order and third-order IFCs, we calculated the phonon transport properties by solving the full phonon Boltzmann transport equation using an iterative method, with a 15×15×15 $q$-point mesh for obtaining the convergent values. The details of the convergence tests for the $q$-point mesh and force interaction cutoff distance are presented in the Supporting Information.

The solved accumulation function at length $x$ indicates the percentage of the lattice thermal conductivity contributed by phonons with MFPs shorter than $x$.[59] In Figure 3a, the computed in-plane accumulation function (red line) is compared with those obtained by fitting the experimental data on $k_{L,\parallel}$[28] using a formula with widely used phonon scattering mechanisms (pink line), and first-principles calculations by Hellman and Broido (blue line).[8] For data fitting, only three acoustic branches are considered, whereas both acoustic and optical branches are included in first-principles calculations. Some discrepancies are anticipated due to the difference in considered phonon branches and employed expressions for phonon scattering rates in the data fitting.

Although the data fitting is usually based on a simplified model, reasonably good agreements between the fitted and measured phonon MFP distributions can be found for materials such as GaN.[73] Here the fitted temperature-dependent $k_{L,\parallel}$ given by measurements ($k_{L,\parallel} \approx 1.72$ W/m·K at 300 K)[28] agrees well with our first-principles-calculated $k_{L,\parallel} \approx 1.71$ W/m·K at room temperature. It should be noted that our calculated room-temperature $k_{L,\parallel}$ value is higher than $k_{L,\parallel}$ of 1.26 –1.3 W/m·K suggested by Hellman and Broido[8] and molecular dynamics (MD) simulations,[7] but lower than $k_{L,\parallel} \approx 2.4$ W/m·K given by other MD simulations that employ interatomic potentials determined by fitting the energy surface from the *ab initio* calculations.[74] Details for the fitted phonon MFPs are given in the Supporting Information. Here the Debye model and our first-principles calculations suggest 50% of the room-temperature $k_{L,\parallel}$ is contributed by phonons with MFPs less than 12 nm and 15 nm, respectively. In comparison, Hellman and Broido also suggest that around 50% of the room-temperature $k_{L,\parallel}$ comes from phonons having MFPs smaller than 8.7 nm.[8] In addition, the fitted accumulation function for direction-averaged MFPs within the $Bi_{0.5}Sb_{1.5}Te_3$ bulk alloy with 20-μm grain sizes[44] is also plotted, where an approximately isotropic $k_L$ is around 0.73 W/m·K at 300 K.

Figure 3b further compares the computed $k_{L,\perp}$ accumulation function with that given by time-domain normal-mode analyses (TD-NMA) for MD simulations by Wang et al.[6] that suggests an overestimated room-temperature $k_{L,\perp} \approx 0.93$ W/m·K by approximating the first Brillouin zone as a cylindrical disk. According to this early study, about 80% of $k_{L,\perp}$ is contributed by phonons with MFPs shorter than 10 nm. In contrast, our first-principles calculations suggest a broader range of phonon MFPs up to ~100 nm but a lower $k_{L,\perp} \approx 0.61$ W/m·K at room temperature. Similar to the treatment of superlattices,[75] the cross-plane phonon transport of bulk $Bi_2Te_3$ has also been modelled analytically,[76, 77] where a *c*-axis interfacial phonon transmissivity for vdW interfaces is given by an acoustic mismatch model.[78] Other MD studies are also available for the impact of vdW interactions on the phonon transport.[79] Besides modeling and simulations, the phonon MFP distribution along the cross-plane direction can be extracted from the measured $k_{L,\perp}$ values for samples with varied thicknesses covering the whole phonon MFP range.[80] However, it is challenging to carry out this inverse phonon transport analysis here because the smallest film thickness measured (>20 nm) is still much longer than majority phonon MFPs. For the in-plane phonon transport, nanoporous films with varied feature sizes can be measured to extract the in-plane phonon MFP distribution.[81] More experimental studies should be performed to compare to our predictions in this work.

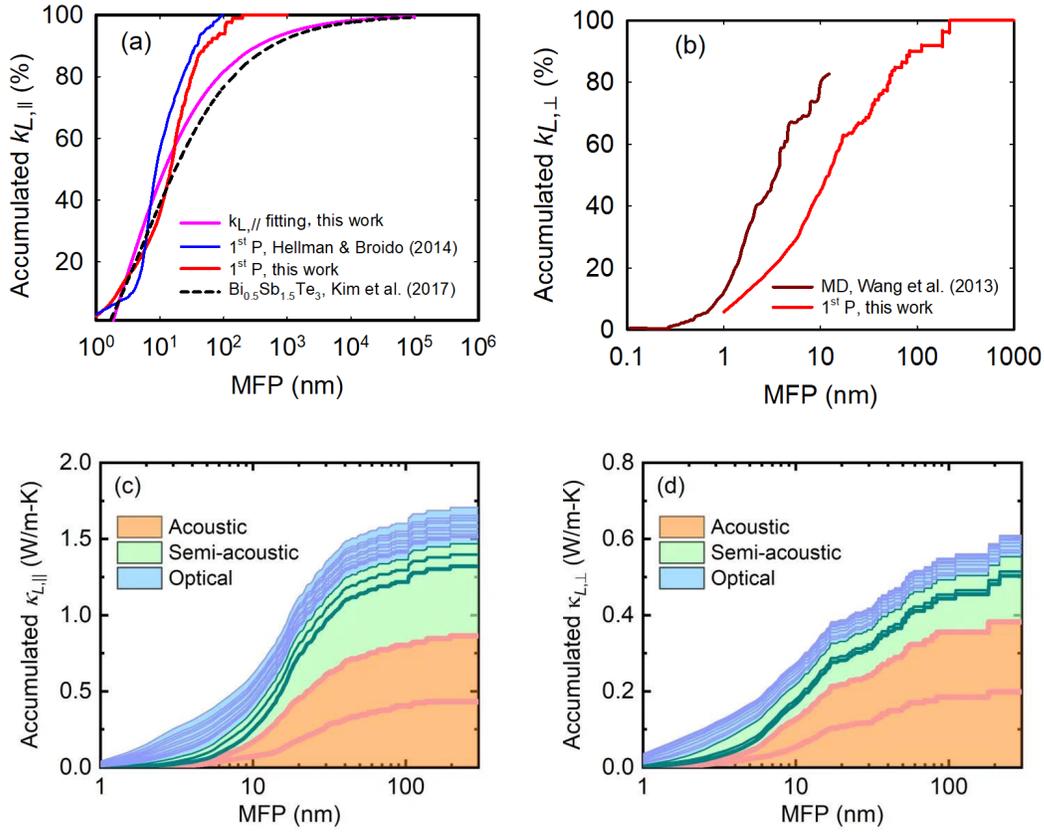

Figure 3. (a) Room-temperature $k_{L,\parallel}$ accumulation function of bulk $Bi_2Te_3$, in comparison with first-principles (marked as "1$^{st}$ P") calculations by Hellman and Broido (blue line).[8] The accumulation function obtained by fitting the experimental data on $k_{L,\parallel}$[28] using a formula with widely used phonon scattering mechanisms is also shown here. The accumulation function for direction-averaged MFPs within the $Bi_{0.5}Sb_{1.5}Te_3$ bulk alloy[44] is plotted in comparison. (b) Room-temperature $k_{L,\perp}$ accumulation function compared with that given by TD-NMA for MD simulations.[6] (c) Room-temperature $k_{L,\parallel}$ accumulation function of bulk $Bi_2Te_3$ decomposed for three ranges of phonon branches. (d) Room-temperature $k_{L,\perp}$ accumulation function of bulk $Bi_2Te_3$ decomposed for three ranges of phonon branches.

For further detailed phonon transport analysis, the contribution from different phonon branches should also be distinguished. Figures 3c and 3d present the thermal conductivity accumulation functions along the in- and cross-plane directions, where contributions from new branches are added from the bottom to the top curves. The bottom three branches are for acoustic modes. Hellman and Broido estimated that around 50% of the room-temperature $k_{L,\parallel}$ for bulk $Bi_2Te_3$ is contributed by the longitudinal acoustic (LA) branch and its intertwined two lowest optical branches within the 1–2 THz frequency range, grouped as "semi-acoustic" branches.[8] At room temperature, our first-principles calculations suggest 35% of $k_{L,\parallel}$ and 28% of $k_{L,\perp}$ is from the LA branch and the two lowest lying optical branches, whereas optical branches in total contribute around 22% of $k_{L,\parallel}$ and 17% of $k_{L,\perp}$.

*Understanding the lattice thermal conductivity of nano- to micro-grained $Bi_2Te_3$.* For both in-plane and cross-plane directions, first-principles-computed accumulation functions in Figure 3 suggest very weak classical phonon size effects within micro-grained $Bi_2Te_3$ samples. In this situation, the experimentally observed lattice thermal conductivity reduction should be mainly attributed to the limited phonon transmission and thus a high $R_K$ across GBs.

Above room temperature, completely diffusive phonon scattering can usually be assumed at GBs, i.e., $P_{GB} \approx 0$.[46, 82] Under this assumption, the lattice thermal conductivity $k_L$ is given by an EMF (see Supporting Information):[18-20]

$$k_L = \sum_p \int_0^{\omega_{max,p}} \frac{c_p(\omega) v_{g,p}(\omega) \Lambda_{eff,p}(\omega)/3}{1 + \frac{2\Lambda_{eff,p}(\omega)}{3T_{GB}(p,\omega)d}} d\omega, \tag{1}$$

where $d$ is the grain size, $p$ indicates the phonon branch, $\omega$ is the phonon angular frequency, $\omega_{max,p}$ is the maximum $\omega$ value for branch $p$, $c_p(\omega)$ is the differential volumetric phonon specific heat, and $v_{g,p}(\omega)$ is the phonon group velocity. Again, $\Lambda_{eff,p}(\omega) = \left[1/\Lambda_{bulk,p}(\omega) + 1/d\right]^{-1}$ is used as the effective phonon MFPs within each grain. Compared with other recent analytical models incorporating the $R_K$ influence,[10, 15, 83-88] Eq. (1) is the only formulation rigorously validated by frequency-dependent phonon Monte Carlo simulations that solve the phonon Boltzmann transport equation for the exact 3D polycrystalline structure. It can also be extended to cryogenic temperatures at which the specularity $P_{GB}$ is no longer negligible.[19] In Eq. (1), the required $T_{GB}(p,\omega)$ generally decreases with an increased $\omega$ value.[10, 89, 90] This is mainly due to stronger GB scattering at increased phonon frequencies. Considering phonon diffraction by dislocations at a GB, analytical models can be found for the phonon scattering rates.[91-94] In a different way, the dislocation-phonon interaction was also modelled for a fully quantized dislocation field, known as a "dislon".[95] Particular attention should also be paid to the phonon transport suppression by the strain field and point defects at a GB. The existing $T_{GB}(p,\omega)$ studies are summarized in a review.[96] Here the DMM suggests an averaged $T_{GB} \approx 0.5$ for a GB with the same material on both sides of the GB.[4] As a simple model, the DMM does not consider the crystal misorientation at a GB and interfacial defects, and fails to explain the imaginary interface limit, i.e., no interfaces and thus 100% phonon transmission.[9] However, the DMM is still a reasonable approximation when detailed $T_{GB}(p,\omega)$ expressions are unavailable. Using such an averaged $T_{GB}$, Eq. (1) combined with the phonon MFP spectrum can be used to estimate $k_L$ values. The accumulation functions of $k_{L,\perp}$ and $k_{L,\parallel}$ can be input into Eq. (1) to compute the lower bound $\langle k_{L,\perp}\rangle$ and upper bound $\langle k_{L,\parallel}\rangle$ of the lattice thermal conductivity, respectively. These two bounds can then be directionally averaged to obtain the effective lattice thermal conductivity $\langle k_L\rangle$ of a polycrystalline material with random grain orientations under a "correlation approximation":[97]

$$\frac{\langle k_L\rangle}{k_{char}} = \frac{1}{3}r^{2/3} + \left[\frac{2}{3} - \frac{2}{9}\frac{(r-1)^2}{r+2}\right]r^{-1/3}, \qquad (2)$$

with $r = \langle k_{L,\perp}\rangle/\langle k_{L,\parallel}\rangle$, $k_{char} = \left(\langle k_{L,\perp}\rangle\langle k_{L,\parallel}\rangle^2\right)^{1/3}$. More discussions on this direction average are given by Yang et al.[98]

Using Eq. (1), grain-size-dependent $\langle k_L\rangle$ is predicted based on first-principles bulk phonon MFPs for $Bi_2Te_3$, assuming an averaged $T_{GB} \approx 0.1 - 0.5$. In Figure 4a, the black $\langle k_L\rangle$ curves only consider the acoustic branches, and the blue $\langle k_L\rangle$ curves consider all phonon branches. Strong GB suppression for optical phonons[99] should be assumed to reduce the discrepancy between the predictions and experimental data, such as micro-grained samples by Kim et al.[39] In addition, the direction-averaged phonon MFPs fitted for a bulk $Bi_{0.5}Sb_{1.5}Te_3$ alloy[44] are also employed to predict $\langle k_L\rangle$ based on Eq. (1), assuming $T_{GB} \approx 0.5$ (dashed pink line). Only acoustic branches are considered in fitted phonon MFPs. The significantly increased point-defect scattering in alloys is considered here and more details are given in the Supporting Information. Despite the variation of alloy compositions, it is found that this curve can better explain the data for various nanostructured $Bi_2Te_3$ alloys. More accurate modeling should also consider unintentional defects in real samples,[4]

GB oxidation,[10, 100] interfacial layers at GBs,[47, 101, 102] interfacial strain fields,[46, 103] porosity,[11, 104] phonon scattering by embedded nanoprecipitates and nanodots,[3] and phonon dispersion variation due to composition changes.[105, 106] Along this line, some studies can also be found on how to tune the thermal resistance for general interfaces.[107, 108]

The suppressed optical phonon contribution due to GBs may be in contrast with the increased importance of optical phonons in nanostructures with strong classical size effects due to phonon boundary scattering.[109] For nanostructures with boundary scattering of phonons, the characteristic length $L_C$ does not have any energy dependence.[110] In this case, optical phonons are less affected for their already short MFPs but acoustic phonons can be dramatically reduced for their MFPs and thus the $k_L$ contribution. In contrast, the GBs as a low-pass filter for phonons can instead decrease the impact of optical phonons on the overall thermal conductivity.

For $Bi_2Te_3$, hot-pressed samples often have preferred grain orientations that invalidate the computed isotropic $\langle k_L \rangle$. For instance, flakes or ribbons often found for synthesized $Bi_2Te_3$-based structures tend to be aligned perpendicular to the pressure direction so that the cross-plane thermal conductivity of a hot-pressed disk can be lower than that for the in-plane direction.[2, 43, 51, 111] In principle, $k_L$ perpendicular to the press direction (open symbols in Figure 4b) should be higher than the isotropic $\langle k_L \rangle$, whereas $k_L$ along the press direction (filled symbols in Figure 4b) is lower than $\langle k_L \rangle$. Some experimental data are not included because the exact grain size is not given.[51, 112] In Figure 4b, the fitted bulk phonon MFPs for $Bi_{0.5}Sb_{1.5}Te_3$[44] assuming $T_{GB} \approx 0.5$ (dashed pink line) can explain the experimental data by Yan et al.[2] Comparison with the data by Han et al.[41] suggests that optical phonons should be largely suppressed for their $\langle k_L \rangle$ contribution. When optical phonons are considered, even a relatively low $T_{GB} \approx 0.25$ still leads to an overpredicted $k_L$ (blue line in Figure 4b).

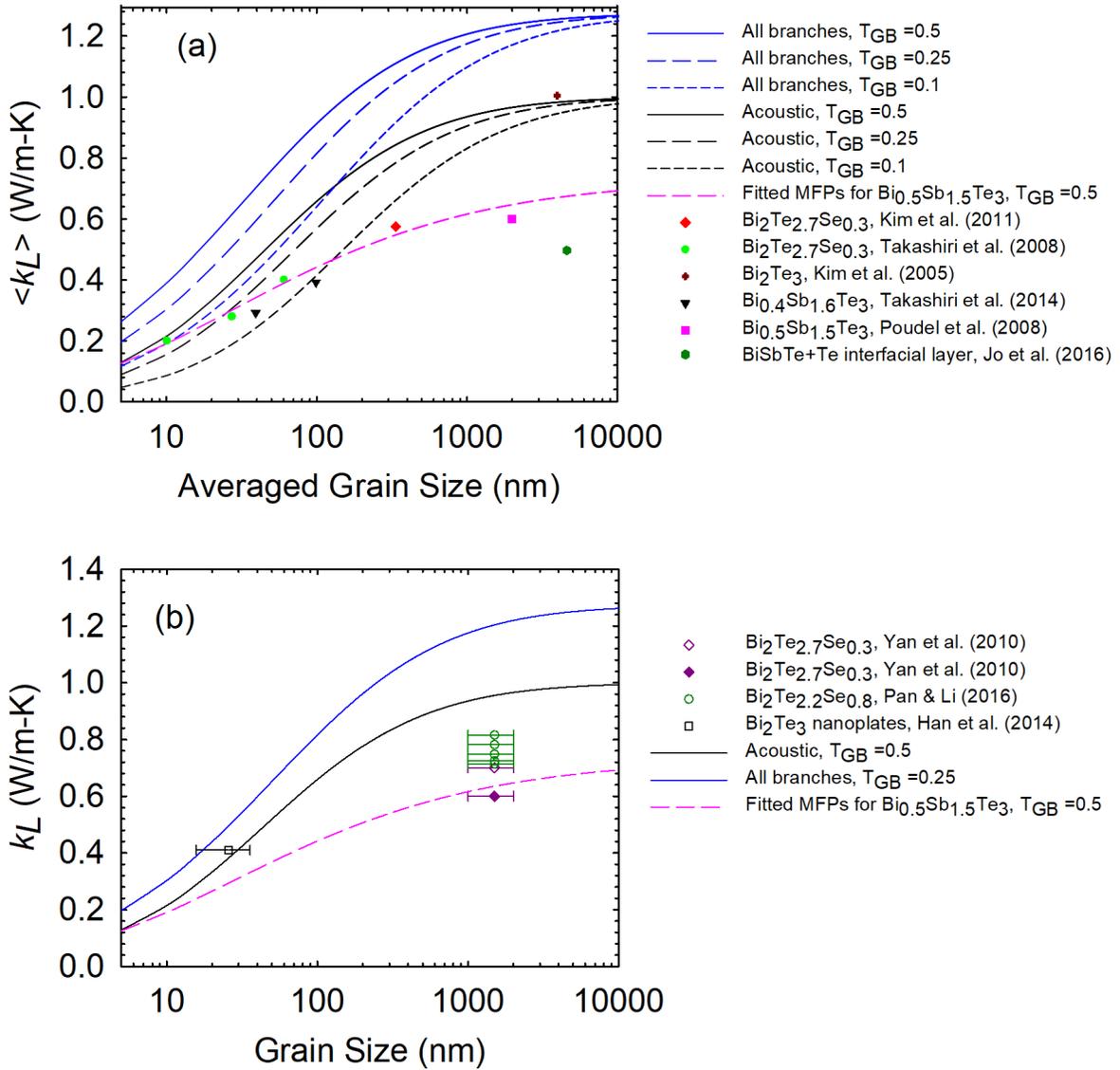

Figure 4. (a) Predicted room-temperature $\langle k_L \rangle$ values as a function of the averaged grain size, in comparison to representative experimental results on polycrystalline $Bi_2Te_3$-based samples. The experimental data are for bulk $Bi_2Te_{2.7}Se_{0.3}$ samples by Kim et al.,[38] bulk $Bi_2Te_3$ samples by Kim et al.,[39] $Bi_2Te_{2.7}Se_{0.3}$ thin films by Takashiri et al.,[13] strained $Bi_{0.4}Sb_{1.6}Te_3$ thin films by Takashiri et al.,[12] nanostructured $Bi_{0.5}Sb_{1.5}Te_3$ bulk alloys by Poudel et al.,[3] and bulk BiSbTe with a 333-nm-thick tellurium interfacial layer at GBs by Jo et al.[40] For $Bi_{0.5}Sb_{1.5}Te_3$ bulk alloys by Poudel et al.,[3] an averaged grain size $d_{eff}$ (see Supporting Information) is used. The averaged grain size and tellurium layer thickness for samples by Jo et al.[40] are approximated as the values for a comparable thin film. (b) Comparison between predicted $\langle k_L \rangle$ and the experimental data for bulk samples with anisotropic lattice thermal conductivities.[41-43] By matching the volumetric interfacial area with that for aligned cubes,[18, 113] an averaged $d \approx 26$ nm for Han et al.[41] is estimated for the initial $Bi_2Te_3$ nanoplates because no remarkable grain growth is found in the transmission electron microscopy image for the bulk samples.

*Computed $R_K$ based on the DMM.* Assuming $T_{GB} \approx 0.5$ and elastic GB phonon scattering, $R_K$ was computed based on the DMM and the full exact phonon dispersion solved from the *ab initio* calculations.[114] $R_K$ based on different crystallographic directions were calculated. The DMM predicted $R_K$ between *c-c*, *a-a*, and *a-c* interfaces are $7.7 \times 10^{-8}$ K·m$^2$/W, $4.2 \times 10^{-8}$ K·m$^2$/W, and $8.4 \times 10^{-8}$ K·m$^2$/W, respectively. The averaged value is $\sim 6.8 \times 10^{-8}$ K·m$^2$/W. The large interfacial resistance originates from the low group velocities and low frequencies of phonon modes in $Bi_2Te_3$. This resistance value is consistent with $R_K \sim 10^{-9} - 10^{-8}$ K·m$^2$/W for nearly perfect interfaces[45] and recent measurements on bonded film-wafer interfaces to represent a twist GB.[46] Figure 5 shows the accumulated interfacial thermal resistance as a function of the phonon frequency. Without optical phonons, the DMM-predicted $R_K$ is as high as $\sim 2 \times 10^{-7}$ K·m$^2$/W. Above values are comparable to the estimated $R_K = 4.9 \times 10^{-8} - 1.15 \times 10^{-7}$ K·m$^2$/W for SiGe alloys, which is obtained from the EMF analysis of the experimental data on microcrystalline samples.[15]

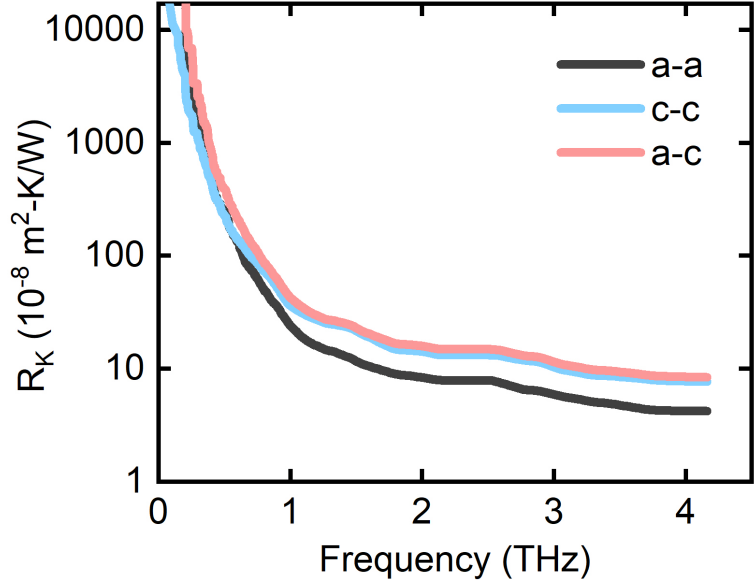

Figure 5. The accumulated interfacial thermal conductance as a function of phonon frequency of interfaces formed by Bi$_2$Te$_3$ crystals with different crystallographic directions: *c-c, a-a,* and *a-c*.

In conclusion, first-principles phonon MFP distributions for both the in-plane and cross-plane directions are re-computed for the bulk Bi$_2$Te$_3$ and split for branch contributions. The cross-plane phonon spectrum is validated with systematic thermal measurements on single-crystal Bi$_2$Te$_3$ nanoplates with various thicknesses. Using these first-principles phonon MFPs to analyze existing data on nano- to micro-grained Bi$_2$Te$_3$ samples, it is found that a high $R_K$ at GBs should be the main cause for the observed significant thermal conductivity reduction and optical phonon contributions should be significantly suppressed. For thermoelectric applications, complicated materials with a large number $n$ of atoms per primitive cell and thus increased $3(n-1)$ optical branches are preferred because more optical branches with slow phonon group velocities lead to a low $k_L$.[115, 116] By neglecting the optical phonon contribution, $k_L \sim n^{-\frac{2}{3}}$ was suggested by Slack for bulk materials[62] and $k_L \sim n^{-1}$ was suggested for acoustic phonons with strong phonon boundary scattering in nanostructures.[115] Further considering optical phonons yields a weaker $k_L \sim n^{-0.4}$ dependence is found for representative oxides.[117] Different from nanostructured boundaries, interfaces such as GBs can help to cut off the percentage of optical-phonon-contributed $k_L$ and benefit complicated materials for thermoelectric and other energy-related applications. The importance of the GBs should be generally acknowledged for nano- to micro-structured bulk tellurides and selenides with short bulk phonon MFPs.[118]

Acknowledgements

A.N., J.H., K.K., J.L. acknowledge the support from NC State Faculty Research and Professional Development Fund. Q. H. acknowledges the support from Craig M. Berge Dean's Fellowship.

References

1. Y. Ma, Q. Hao, B. Poudel, Y. Lan, B. Yu, D. Wang, G. Chen and Z. Ren, Nano Letters **8** (8), 2580-2584 (2008).
2. X. Yan, B. Poudel, Y. Ma, W. S. Liu, G. Joshi, H. Wang, Y. Lan, D. Wang, G. Chen and Z. F. Ren, Nano Letters **10** (9), 3373-3378 (2010).
3. B. Poudel, Q. Hao, Y. Ma, Y. Lan, A. Minnich, B. Yu, X. Yan, D. Wang, A. Muto, D. Vashaee, X. Chen, J. Liu, M. S. Dresselhaus, G. Chen and Z. Ren, Science **320** (5876), 634-638 (2008).
4. Q. Hao, G. Zhu, G. Joshi, X. Wang, A. Minnich, Z. Ren and G. Chen, Applied Physics Letters **97** (6), 063109 (2010).
5. S. K. Bux, R. G. Blair, P. K. Gogna, H. Lee, G. Chen, M. S. Dresselhaus, R. B. Kaner and J.-P. Fleurial, Advanced Functional Materials **19** (15), 2445-2452 (2009).
6. Y. Wang, B. Qiu, A. JH McGaughey, X. Ruan and X. Xu, Journal of heat transfer **135** (9) (2013).
7. B. Qiu and X. Ruan, Physical Review B **80** (16), 165203 (2009).
8. O. Hellman and D. A. Broido, Physical Review B **90** (13), 134309 (2014).
9. G. Chen, *Nanoscale Energy Transport and Conversion: A Parallel Treatment of Electrons, Molecules, Phonons, and Photons*. (Oxford University Press, New York, 2005).
10. Z. Wang, J. E. Alaniz, W. Jang, J. E. Garay and C. Dames, Nano Letters **11** (6), 2206-2213 (2011).
11. M. Takashiri, S. Tanaka, H. Hagino and K. Miyazaki, Journal of Applied Physics **112** (8), 084315 (2012).

12. M. Takashiri, S. Tanaka, H. Hagino and K. Miyazaki, International Journal of Heat and Mass Transfer **76**, 376-384 (2014).
13. M. Takashiri, K. Miyazaki, S. Tanaka, J. Kurosaki, D. Nagai and H. Tsukamoto, Journal of Applied Physics **104** (8), 084302 (2008).
14. X. Wu, J. Lee, V. Varshney, J. L. Wohlwend, A. K. Roy and T. Luo, Scientific Reports **6** (1), 22504 (2016).
15. C.-W. Nan and R. Birringer, Physical Review B **57** (14), 8264 (1998).
16. P. L. Kapitza, J. Phys. (Moscow) **4**, 181 (1941).
17. E. T. Swartz and R. O. Pohl, Reviews of Modern Physics **61** (3), 605-668 (1989).
18. Q. Hao, Journal of Applied Physics **111** (1), 014307 (2012).
19. Q. Hao, Journal of Applied Physics **116** (3), 034305 (2014).
20. Q. Hao, H. Zhao, Y. Xiao and D. Xu, Journal of Applied Physics **123** (1), 014303 (2018).
21. M. G. Muraleedharan, K. Gordiz, A. Rohskopf, S. T. Wyant, Z. Cheng, S. Graham and A. Henry, arXiv preprint arXiv:2011.01070 (2020).
22. J. Hickman and Y. Mishin, Physical Review Materials **4** (3), 033405 (2020).
23. C. Li and Z. Tian, Frontiers in Physics **7** (3) (2019).
24. S. Fujii, T. Yokoi, C. A. J. Fisher, H. Moriwake and M. Yoshiya, Nat Commun **11** (1), 1854-1854 (2020).
25. F. Rosi, B. Abeles and R. Jensen, Journal of Physics and Chemistry of Solids **10** (2-3), 191-200 (1959).
26. H. Goldsmid, Journal of Applied Physics **32** (10), 2198-2202 (1961).
27. J. Fleurial, L. Gailliard, R. Triboulet, H. Scherrer and S. Scherrer, Journal of Physics and Chemistry of Solids **49** (10), 1237-1247 (1988).
28. C. Satterthwaite and R. Ure Jr, Physical review **108** (5), 1164 (1957).
29. C. Champness, W. Muir and P. Chiang, Canadian Journal of Physics **45** (11), 3611-3626 (1967).
30. H.-W. Jeon, H.-P. Ha, D.-B. Hyun and J.-D. Shim, Journal of Physics and Chemistry of Solids **52** (4), 579-585 (1991).
31. A. Jacquot, N. Farag, M. Jaegle, M. Bobeth, J. Schmidt, D. Ebling and H. Böttner, Journal of Electronic Materials **39** (9), 1861-1868 (2010).
32. H. Goldsmid, Proceedings of the Physical Society. Section B **69** (2), 203 (1956).
33. Z. Tian, K. Esfarjani, J. Shiomi and A. S. Henry, Applied Physics Letters **99** (5), 141 (2011).
34. D. Li, Y. Wu, P. Kim, L. Shi, P. Yang and A. Majumdar, Applied Physics Letters **83** (14), 2934-2936 (2003).
35. Y. Xiao, Q. Chen, D. Ma, N. Yang and Q. Hao, ES Materials & Manufacturing **5**, 2–18 (2019).
36. A. M. Marconnet, M. Asheghi and K. E. Goodson, Journal of heat transfer **135** (6) (2013).
37. D. Kong, W. Dang, J. J. Cha, H. Li, S. Meister, H. Peng, Z. Liu and Y. Cui, Nano letters **10** (6), 2245-2250 (2010).
38. C. Kim, D. H. Kim, Y. S. Han, J. S. Chung, S. Park, S. Park and H. Kim, Materials Research Bulletin **46** (3), 407-412 (2011).
39. D.-H. Kim and T. Mitani, Journal of Alloys and Compounds **399** (1), 14-19 (2005).
40. S. Jo, S. H. Park, H. W. Ban, D. H. Gu, B.-S. Kim, J. H. Son, H.-K. Hong, Z. Lee, H.-S. Han, W. Jo, J. E. Lee and J. S. Son, Journal of Alloys and Compounds **689**, 899-907 (2016).

41. G. Han, Z.-G. Chen, L. Yang, M. Hong, J. Drennan and J. Zou, ACS applied materials & interfaces **7** (1), 989-995 (2015).
42. X. Yan, B. Poudel, Y. Ma, W. Liu, G. Joshi, H. Wang, Y. Lan, D. Wang, G. Chen and Z. Ren, Nano letters **10** (9), 3373-3378 (2010).
43. Y. Pan and J.-F. Li, NPG Asia Materials **8** (6), e275-e275 (2016).
44. H.-S. Kim, S. I. Kim, K. H. Lee, S. W. Kim and G. J. Snyder, physica status solidi (b) **254** (5), 1600103 (2017).
45. R. M. Costescu, M. A. Wall and D. G. Cahill, Physical Review B **67** (5), 054302 (2003).
46. D. Xu, R. Hanus, Y. Xiao, S. Wang, G. J. Snyder and Q. Hao, Materials Today Physics **6**, 53-59 (2018).
47. S. I. Kim, K. H. Lee, H. A. Mun, H. S. Kim, S. W. Hwang, J. W. Roh, D. J. Yang, W. H. Shin, X. S. Li and Y. H. Lee, Science **348** (6230), 109-114 (2015).
48. K.-C. Kim, J. Lee, B. K. Kim, W. Y. Choi, H. J. Chang, S. O. Won, B. Kwon, S. K. Kim, D.-B. Hyun, H. J. Kim, H. C. Koo, J.-H. Choi, D.-I. Kim, J.-S. Kim and S.-H. Baek, Nat Commun **7** (1), 12449 (2016).
49. J.-Y. Hwang, J. Kim, H.-S. Kim, S.-I. Kim, K. H. Lee and S. W. Kim, Advanced Energy Materials **8** (20), 1800065 (2018).
50. J. Mao, Y. Wang, Z. Liu, B. Ge and Z. Ren, Nano Energy **32**, 174-179 (2017).
51. R. Deng, X. Su, Z. Zheng, W. Liu, Y. Yan, Q. Zhang, V. P. Dravid, C. Uher, M. G. Kanatzidis and X. Tang, Science Advances **4** (6), eaar5606 (2018).
52. X. Lu, Q. Hao, M. Cen, G. Zhang, J. Sun, L. Mao, T. Cao, C. Zhou, P. Jiang and X. Yang, Nano letters **18** (5), 2879-2884 (2018).
53. X. Lu, O. Khatib, X. Du, J. Duan, W. Wei, X. Liu, H. A. Bechtel, F. D'Apuzzo, M. Yan, A. Buyanin, Q. Fu, J. Chen, M. Salmeron, J. Zeng, M. B. Raschke, P. Jiang and X. Bao, Advanced Electronic Materials **4** (1), 1700377 (2018).
54. M. Chen, R. C. Haddon, R. Yan and E. Bekyarova, Materials Horizons **4** (6), 1054-1063 (2017).
55. S. Wang, D. Xu, R. Gurunathan, G. J. Snyder and Q. Hao, Journal of Materiomics (2020).
56. J. Liu, J. Zhu, M. Tian, X. Gu, A. Schmidt and R. Yang, Review of Scientific Instruments **84** (3), 034902 (2013).
57. D. G. Cahill, P. V. Braun, G. Chen, D. R. Clarke, S. Fan, K. E. Goodson, P. Keblinski, W. P. King, G. D. Mahan, A. Majumdar, H. J. Maris, S. R. Phillpot, E. Pop and L. Shi, Applied Physics Reviews **1** (1), 011305 (2014).
58. P. Jiang, X. Qian and R. Yang, Journal of Applied Physics **124** (16), 161103 (2018).
59. F. Yang and C. Dames, Physical Review B **87** (3), 035437 (2013).
60. C. Hua and A. J. Minnich, Journal of Applied Physics **117** (17), 175306 (2015).
61. M. T. Pettes, J. Maassen, I. Jo, M. S. Lundstrom and L. Shi, Nano Letters **13** (11), 5316-5322 (2013).
62. D. Park, S. Park, K. Jeong, H.-S. Jeong, J. Y. Song and M. H. Cho, Scientific Reports **6** (1), 19132 (2016).
63. J. P. Fleurial, L. Gailliard, R. Triboulet, H. Scherrer and S. Scherrer, Journal of Physics and Chemistry of Solids **49** (10), 1237-1247 (1988).
64. C. B. Satterthwaite and R. W. Ure, Physical Review **108** (5), 1164-1170 (1957).
65. B. Qiu and X. Ruan, Applied Physics Letters **97** (18), 183107 (2010).
66. G. Kresse and J. Hafner, Physical review B **47** (1), 558 (1993).

67. G. Kresse and J. Furthmüller, Physical Review B **54** (16), 11169-11186 (1996).
68. D. S. Kim, O. Hellman, J. Herriman, H. L. Smith, J. Y. Y. Lin, N. Shulumba, J. L. Niedziela, C. W. Li, D. L. Abernathy and B. Fultz, Proceedings of the National Academy of Sciences **115** (9), 1992-1997 (2018).
69. O. Hellman, P. Steneteg, I. A. Abrikosov and S. I. Simak, Physical Review B **87** (10), 104111 (2013).
70. D. M. Ceperley and B. J. Alder, Physical review letters **45** (7), 566 (1980).
71. M. J. Mehl, D. Hicks, C. Toher, O. Levy, R. M. Hanson, G. Hart and S. Curtarolo, Computational Materials Science **136**, S1-S828 (2017).
72. D. Hicks, M. J. Mehl, E. Gossett, C. Toher, O. Levy, R. M. Hanson, G. Hart and S. Curtarolo, Computational Materials Science **161**, S1-S1011 (2019).
73. Q. Hao, H. Zhao and Y. Xiao, Journal of Applied Physics **121** (20), 204501 (2017).
74. B.-L. Huang and M. Kaviany, Physical Review B **77** (12), 125209 (2008).
75. C. Dames and G. Chen, Journal of Applied Physics **95** (2), 682-693 (2004).
76. P. Al-Alam, G. Pernot, M. Isaiev, D. Lacroix, M. De Vos, N. Stein, D. Osenberg and L. Philippe, Physical Review B **100** (11), 115304 (2019).
77. K. H. Park, M. Mohamed, Z. Aksamija and U. Ravaioli, Journal of Applied Physics **117** (1), 015103 (2015).
78. R. Prasher, Applied Physics Letters **94** (4), 041905 (2009).
79. X. Yu, D. Ma, C. Deng, X. Wan, M. An, H. Meng, X. Li, X. Huang and N. Yang, Chinese Physics Letters **38** (1), 014401 (2021).
80. H. Zhang, C. Hua, D. Ding and A. J. Minnich, Scientific Reports **5** (1), 9121 (2015).
81. Q. Hao, Y. Xiao and Q. Chen, Materials Today Physics **10**, 100126 (2019).
82. X. Wang and B. Huang, Scientific reports **4** (2014).
83. H.-S. Yang, G.-R. Bai, L. Thompson and J. Eastman, Acta Materialia **50** (9), 2309-2317 (2002).
84. H. Dong, B. Wen and R. Melnik, Scientific reports **4**, 7037 (2014).
85. D. S. Smith, S. Fayette, S. Grandjean, C. Martin, R. Telle and T. Tonnessen, Journal of the American Ceramic Society **86** (1), 105-111 (2003).
86. H. Dong, B. Wen and R. Melnik, Sci Rep **4** (1), 7037 (2014).
87. F. Badry and K. Ahmed, AIP Advances **10** (10), 105021 (2020).
88. T. Hori, J. Shiomi and C. Dames, Applied Physics Letters **106** (17), 171901 (2015).
89. P. K. Schelling, S. R. Phillpot and P. Keblinski, Journal of Applied Physics **95** (11), 6082-6091 (2004).
90. S.-H. Ju and X.-G. Liang, Journal of Applied Physics **113** (5), 053513-053517 (2013).
91. R. Hanus, A. Garg and G. J. Snyder, Communications Physics **1** (1), 1-11 (2018).
92. P. Klemens, Proceedings of the Physical Society. Section A **68** (12), 1113 (1955).
93. M. Omini and A. Sparavigna, Physical Review B **61** (10), 6677 (2000).
94. P. Carruthers, Physical Review **114** (4), 995 (1959).
95. M. Li, Z. Ding, Q. Meng, J. Zhou, Y. Zhu, H. Liu, M. S. Dresselhaus and G. Chen, Nano Letters **17** (3), 1587-1594 (2017).
96. Q. Hao and J. Garg, ES Materials & Manufacturing **14**, 36-50 (2021).
97. E. A. Mityushov, R. A. Adamesku and P. V. Gel'd, Journal of engineering physics **47** (3), 1052-1056 (1984).
98. F. Yang, T. Ikeda, G. J. Snyder and C. Dames, Journal of Applied Physics **108** (3), 034310 (2010).

99. Z. Zheng, X. Chen, B. Deng, A. Chernatynskiy, S. Yang, L. Xiong and Y. Chen, Journal of Applied Physics **116** (7), 073706 (2014).
100. S. Wang, D. Xu, R. Gurunathan, G. J. Snyder and Q. Hao, Journal of Materiomics **6** (2), 248-255 (2020).
101. S. K. Bux, R. G. Blair, P. K. Gogna, H. Lee, G. Chen, M. S. Dresselhaus, R. B. Kaner and J. P. Fleurial, Advanced Functional Materials **19** (15), 2445-2452 (2009).
102. Y. Lan, B. Poudel, Y. Ma, D. Wang, M. S. Dresselhaus, G. Chen and Z. Ren, Nano letters **9** (4), 1419-1422 (2009).
103. R. Hanus, A. Garg and G. J. Snyder, Communications Physics **1** (1), 78 (2018).
104. Q. Hao, G. J. Coleman, D. Xu, E. R. Segal, P. Agee, S. Wu and P. Lucas, Frontiers in Energy Research **6** (21) (2018).
105. K. Park, K. Ahn, J. Cha, S. Lee, S. I. Chae, S.-P. Cho, S. Ryee, J. Im, J. Lee and S.-D. Park, Journal of the American Chemical Society **138** (43), 14458-14468 (2016).
106. F. Kargar, E. H. Penilla, E. Aytan, J. S. Lewis, J. E. Garay and A. A. Balandin, Applied Physics Letters **112** (19), 191902 (2018).
107. L. Yang, X. Wan, D. Ma, Y. Jiang and N. Yang, Physical Review B **103** (15), 155305 (2021).
108. S. Deng, C. Xiao, J. Yuan, D. Ma, J. Li, N. Yang and H. He, Applied Physics Letters **115** (10), 101603 (2019).
109. Z. Tian, K. Esfarjani, J. Shiomi, A. S. Henry and G. Chen, Applied Physics Letters **99** (5), 053122 (2011).
110. Y. Xiao, Q. Chen, D. Ma, N. Yang and Q. Hao, ES Materials & Manufacturing **5**, 2-18 (2019).
111. J. Shen, L. Hu, T. Zhu and X. Zhao, Applied Physics Letters **99** (12), 124102 (2011).
112. C. V. Manzano, B. Abad, M. Muñoz Rojo, Y. R. Koh, S. L. Hodson, A. M. Lopez Martinez, X. Xu, A. Shakouri, T. D. Sands, T. Borca-Tasciuc and M. Martin-Gonzalez, Scientific reports **6**, 19129-19129 (2016).
113. Q. Hao, Journal of Applied Physics **111** (1), 014309 (2012).
114. H. Subramanyan, K. Kim, T. Lu, J. Zhou and J. Liu, AIP Advances **9** (11), 115116 (2019).
115. E. S. Toberer, A. Zevalkink and G. J. Snyder, Journal of Materials Chemistry **21** (40), 15843-15852 (2011).
116. S. A. Miller, P. Gorai, B. R. Ortiz, A. Goyal, D. Gao, S. A. Barnett, T. O. Mason, G. J. Snyder, Q. Lv and V. Stevanović, Chemistry of Materials **29** (6), 2494-2501 (2017).
117. Q. Hao, D. Xu, N. Lu and H. Zhao, Physical Review B **93** (20), 205206 (2016).
118. D. Wu, X. Chen, F. Zheng, H. Du, L. Jin and R. E. Dunin-Borkowski, ACS Applied Energy Materials **2** (4), 2392-2397 (2019).